\documentclass[twocolumn, preprintnumbers,amsmath,amssymb]{revtex4}

\usepackage[dvips]{graphicx}
\usepackage{dcolumn}
\usepackage{bm}

\begin{document}

\preprint{}

\title{Mechanism of Laser-induced Field Emission}

\author{Hirofumi Yanagisawa${}^{1,\dagger}$}
\email{hirofumi@phys.ethz.ch}
\author{Matthias Hengsberger${}^{1}$}
\author{Dominik Leuenberger${}^{1}$}
\author{Martin Kl\"{o}ckner${}^{1}$}
\author{Christian Hafner${}^{2}$}
\author{Thomas Greber${}^{1}$}
\author{J\"{u}rg Osterwalder${}^{1}$}
\affiliation{${}^{1}$\mbox{Physik Institut, Universit\"{a}t Z\"{u}rich, Winterthurerstrasse 190, CH-8057 Z\"{u}rich, Switzerland }\\
${}^2$\mbox{Laboratory for Electromagnetic Fields and Microwave Electronics, ETH Z\"{u}rich, Gloriastrasse 35, CH-8092 Z\"{u}rich, Switzerland} \\
${}^\dagger$\mbox{Present address: Department of Physics, ETH Z\"{u}rich, Wolfgang-Pauli-Strasse 16, CH-8093 Z\"{u}rich, Swizerland}}

\begin{abstract}
We have measured electron energy distribution curves (EDCs) of the laser-induced field emission from a tungsten tip. Field emission from photo-excited nonequilibrium electron distributions were clearly observed, while no enhanced field emission due to optical electric fields appeared up to values of 1.3 V/nm. Thus, we experimentally confirm the emission mechanism. Simulated transient EDCs show that electron dynamics plays a significant role in the laser-induced field emission. The results should be useful to find optimal parameters for defining the temporal and spectral characteristics of electron pulses for many applications based on pulsed field emission.
\end{abstract}

\date{\today}
\maketitle

Applying strong electric fields (2-6 V/nm) to a metallic tip with nanometer sharpness enables field emission due to electron tunneling into the vacuum, producing continuous electron beams with high brightness and coherence \cite{gomer93, fursey03, fink86,  nagaoka98,  cho04}. Illumination of such tips by femtosecond laser pulses has realized pulsed field emission with spatio-temporal control with femtosecond and nanometer resolution, making it attractive for both basic research and new applications like time-resolved electron microscopy, spectroscopy, holography, and also free electron lasers \cite{hommelhoff06a,yanagisawa09,yanagisawa10,aeschlimann07}. Despite intensive research over the last half decade \cite{hommelhoff06a,yanagisawa09,yanagisawa10,aeschlimann07,hommelhoff06b,ropers07,ganter08,wu08,tsujino09, hommelhoff10,ropers10}, the emission mechanism is still controversial with the fundamental question being whether the optical fields of the laser pulse interact with the electrons in the tip as particles (photons) or via the electrical field proper. 

When a focused laser pulse illuminates a metallic tip, optical electric fields are enhanced at the tip apex due to plasmonic effects, and the enhanced fields induce pulsed field emission in combination with a moderate DC voltage applied to the tip. Depending on the strength of the enhanced fields, different field emission mechanisms are considered to become dominant \cite{hommelhoff06a,yanagisawa09}. For relatively weak fields, single-electron excitations by single- and multi-photon absorption are prevalent, and photo-excited electrons are tunneling through the surface potential barrier (\emph{photo-field emission}). On the other hand, very strong fields largely modify the tunneling barrier and prompt field emission from the Fermi level (\emph{optical field emission}). 

To reveal the emission mechanisms, measuring electron energy distribution curves (EDCs) is the most direct method. Here we present such data, and we experimentally confirm that the emission mechanism can be quantitatively described within the pure photo-field emission model up to optical fields of 1.3 V/nm. Simulated transient EDCs show that electron dynamics plays a significant role in the laser-induced field emission.

\begin{figure}[!b]
\begin{center}
\vspace{-3pt}
\includegraphics[bb=0 0 566 571, scale=0.37]{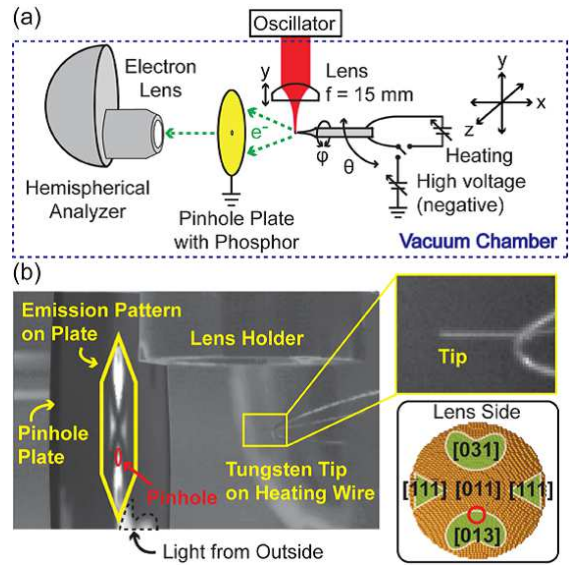} 
\end{center}
\vskip -\lastskip \vskip -3pt
\caption{\label{fig:epsart}
(color online) (a) a schematic diagram of the experimental setup (see text for further description). (b) a photograph of the experimental setup. The inset is the front view of the atomic structure of the tip apex based on a ball model. }
\label{fig:label-1}
\end{figure}

Fig. 1(a) schematically illustrates our experimental setup. A tungsten tip with its axis along the [011] crystal direction is mounted inside a vacuum chamber ($9 \cdot 10^{-11}$ $mbar$). Laser pulses are generated in a Ti:sapphire oscillator (center wave length: 800 nm; repetition rate: 76 MHz). The temporal spread of the laser-pulse intensity profile is estimated to be roughly 100 fs in FWHM just at the tip apex.The laser light was focused to 4 $\mu$m ($1/e^2$ radius) onto the tip apex. Linearly polarized laser light was used, with the polarization vector parallel to the tip axis. The tip can be heated to clean the apex and also negatively biased for field emission. A pinhole plate with phosphor coating was mounted in front of the tip to observe emission patterns from the tip apex, and to define a specific emission site for electon spectroscopy. A hemispherical analyzer (VG: CLAM2) is used to measure EDCs of the emitted electrons passing through the pinhole. The tip can be moved along five axes as used in our previous work \cite{yanagisawa09, yanagisawa10}. The tip axis is set to be orthogonal to the pinhole plate.

A photograph of the experimental setup is shown in Fig. 1(b). The field emission pattern of the clean tungsten tip can be observed on the phosphor plate where the most intense electron emission is observed around the [310]-type facets. The emission sites are highlighted by green areas on the schematic front view of the tip apex in the inset of Fig. 1(b). The pinhole is positioned at the edge of a [310] type facet in Fig. 1(b). The position of the pinhole is roughly indicated by a red circle in the inset; the selected site is the most intense emission site in the laser-induced field emission \cite{yanagisawa09, yanagisawa10}.

\begin{figure}[t!]
\begin{center}

\includegraphics[bb=0 0 408 484, scale=0.55]{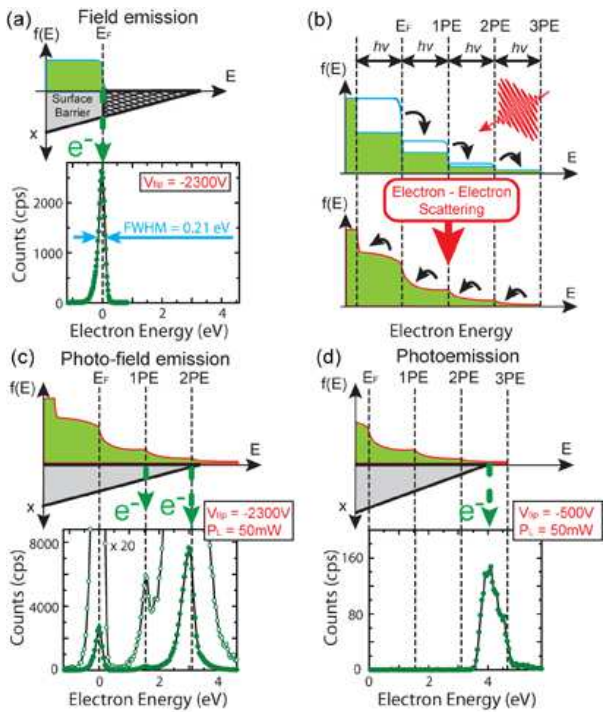} 
\end{center}
\vskip -\lastskip \vskip -3pt
\caption{\label{fig:epsart}
(color online) (a) EDC of field emission from the tungsten tip together with a schematic diagram of field emission from a Fermi-Dirac distribution. The tip voltage $V_{tip}$ was -2300 V. (b) schematic diagrams of nonequiliburium electron distributions just after laser absorption without (upper panel) and with (lower panel) electron-electron scattering. (c) and (d) show experimental EDCs and schematic diagrams of photo-field emission and photoemission, respectively. The magnified spectrum is also shown in (c). (c) $V_{tip}$ = -2300 V and the laser power $P_{L}$ = 50 mW. (d) $V_{tip}$ = -500 V and  $P_{L}$ = 50 mW.}
\vspace{-3pt}
\label{fig:label-2}
\end{figure}

Fig. 2(a) shows an EDC of field emission at a tip voltage $V_{tip}$ of -2300 V. The peak of the spectrum defines the Fermi energy $E_{F}$ at 0 eV. The spectrum shows a typical asymmetric peak, which can be understood by the diagram in Fig. 2(a). The field emission current is influenced by two factors: 1) the electron occupation number and 2) the transmission probability through the surface barrier \cite{gomer93,fursey03,murphy56,young59}. The occupation number is given by an electron distribution function $f(E)$, which is the Fermi-Dirac distribution function in the case of field emission, and the transmission probability depends exponentially on an area of the surface barrier indicated by the hatched area. Therefore, the positive energy side of the spectrum falls off due to a rapid decrease of the occupation number, while the negative energy side falls off because of the exponential decay of the transmission probability due to the increase of the surface barrier area. Thus a
  typical field emission spectrum shows such an asymmetric peak. An energy spread of 0.21 eV was observed, which is close to the value measured with 1 meV energy resolution in previous work (0.19 eV) \cite{ogawa96}, confirming our reasonable energy resolution.

In photo-field emission, the electron distribution is strongly modified by single-electron excitations due to multi-photon absorption, resulting in a nonequilibrium distribution characterized by a steplike profile as illustrated in the upper panel of Fig. 2(b) \cite{wu08,lisowski04,rethfeld02}; the width of each step corresponds to the photon energy $h\nu$ (= 1.55 eV). Here we identify the step edges of one-photon excitation (1PE), 2PE and 3PE as shown in Fig. 2(b). In a real situation, however, the excited electrons relax mainly by electron-electron ({\it e-e}) scattering on a time scale of a few femtoseconds, which is shorter than our laser pulses. As a result, the electron dynamics is reflected in a smeared electron distribution, as shown in the lower panel of Fig. 2(b). This feature should be reflected in the measured EDCs.

Fig. 2(c) shows an EDC of laser-induced field emission at $V_{tip}$ = -2300 V and a laser power $P_{L}$ of 50 mW. The spectrum shows the field emission peak undisturbed with identical shape and intensity as in Fig. 2(a), and additional peaks at the 1PE and 2PE edges are clearly observed. The latter show the same asymmetric shape as the field emission peak. Thus, photo-field emission is confirmed experimentally. Regarding the relative intensities, photo-field emission from 2PE is much higher than that from 1PE even though the occupation number at 1PE is higher; this is because the transmission probability at 2PE is quite high. Note that photo-field emission from 2PE has not been observed for excitation with a CW laser \cite{gao85}; electron distributions are supposed to be largely different from our case. 

An EDC at $V_{tip}$ = -500 V and $P_{L}$ = 50 mW is shown in Fig. 2(d). In this spectrum, the nonequilibrium electron distribution function becomes more noticeable. Field emission and photo-field emission processes are suppressed due to the low DC field, and photoemission over the surface barrier dominates, for which 3PE is required. The peak shows a spectral shape completely different from that of field emission. The peak maximum is located approximately 0.65 eV below the 3PE edge. Since the transmission probability is unity throughout the photoemission regime, the peak shape reflects more closely the electron distribution function. We observe that it is strongly modulated by the {\it e-e} scattering processes.


\begin{figure}[t!]
\begin{center}
\includegraphics[bb=0 0 475 438, scale=0.52]{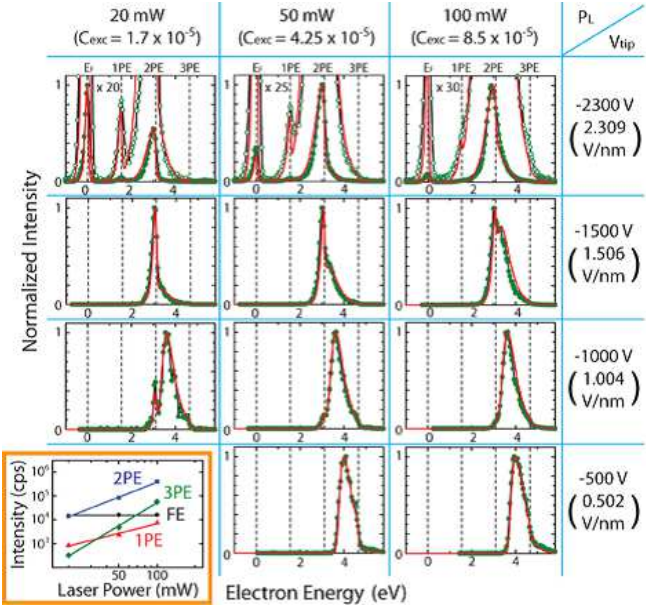}
\end{center}
\vskip -\lastskip \vskip -3pt
\caption{\label{fig:epsart}
(color online) Measured and simulated EDCs for different tip voltages (DC fields $F_{DC}$) and laser powers (excitation constant $C_{exc}$, appears in eq. (1)). All the spectra are normalized at their maximum values. The measured EDCs are shown by green dots with lines, and the simulated EDCs by red lines. Magnified spectra are also shown in the spectra at $V_{tip}$ =-2300 V by green circles with lines for the experiments and red lines for the simulations. The inset (bottom left) shows the measured laser power dependence of intensities for each excitation order ($V_{tip}$=-2300 V). These intensities were obtained by integrating intensities over the respective energy regions: field emission (FE) is for -1.2 eV $\sim$ 0.3 eV, photo-field emission and photoemission from 1PE for 0.3 eV $\sim$ 1.85 eV, 2PE for 1.85 eV $\sim$ 3.4 eV and 3PE for 3.4 eV $\sim$ 4.95 eV. The lines are fitting curves with power functions; exponents are 1.3, 2.1 and 3.2 for 1PE, 2PE and 3PE, respectively. 
}
\vspace{-3pt}
\label{fig:label-3}
\end{figure}


Further investigations of the emission mechanisms were done by systematically measuring EDCs for various tip voltages and laser powers as shown in Fig. 3. All the data indicate pure photo-field emission, but dominant processes change with tip voltage and laser power. As a rule of thumb, with decreasing tip voltage and increasing laser power, electron emission from higher-order photon excitation becomes dominant. These data show a remarkable tunability of the emission processes via these two parameters.

Throughout the whole spectra, no clear onset of optical field emission is observed. The inset of Fig. 3 shows a characteristic increase of intensities as a function of laser power for each excitation order at $V_{tip}$=-2300 V. The field emission (FE) intensity, on the other hand, remains constant, showing no enhanced emission due to optical field emission. Note that laser power values of 20, 50, 100 mW correspond to local optical fields of 0.3, 0.9, 1.3 V/nm on the tip apex (a field enhancement by a factor of 2 was included according to Refs. \cite{yanagisawa09, yanagisawa10}).

\begin{figure}[b]
\begin{center}
\vspace{-3pt}
\includegraphics[bb=0 0 622 429, scale=0.38]{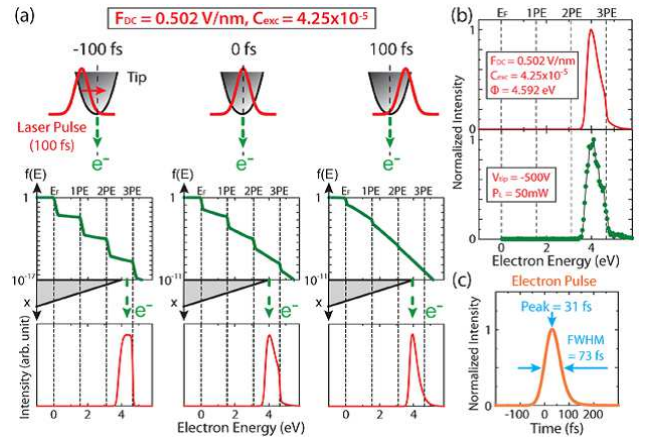} 
\end{center}
\vskip -\lastskip \vskip -3pt
\caption{\label{fig:epsart}
(color online) (a) Schematic diagrams and examples of simulations at three time steps. Transient electron distribution functions are shown by green lines and EDCs by red lines. The three EDCs are not drawn to scale. (b) time-integrated EDCs (upper panel) and experimental EDC (lower panel). (c) temporal line profile of the simulated electron pulse.}
\label{fig:label-4}
\end{figure}

The influence of electron dynamics in photo-field emission was clarified by simulating transient EDCs. The basic approach was the same as that used in Refs. \cite{wu08, rethfeld02, rethfeld99, fatti00}. {\it e-e} interaction, electron-phonon ({\it e-p}) interaction and single electron excitation by single-photon absorption are included in a system of Boltzmann's equations to obtain the temporal evolution of the distribution function of the electron gas and the phonon gas. Fermi-Dirac and Bose-Einstein distributions at room temperature were assumed as initial conditions. The collision terms that describe {\it e-e} and {\it e-p} scattering were given in Refs. \cite{rethfeld02, rethfeld99}, and the term for the electron-gas excitation by the pump pulse was given in Ref. \cite{fatti00} as 
\small{
\begin{eqnarray}
H(E,t) = C_{exc} F(t)_{gauss} \{ \sqrt{E-\hbar\nu}f(E-\hbar\nu)[1-f(E)]\nonumber\\-\sqrt{E+\hbar\nu}f(E)[1-f(E+\hbar\nu)] \}
\end{eqnarray}
}
\normalsize where $F(t)_{gauss}$ is a Gaussian function with a maximum of unity and a temporal spread of 100 fs, and $C_{exc}$ is an excitation constant which is proportional to the laser power and is used as a fitting parameter in our simulations. To calculate {\it e-e} and {\it e-p} scattering, we used free-electron and Debye models with a Fermi energy of 9.2 eV \cite{islam09}, a Debye temperature of 400 K \cite{kittel}, and a sound speed for longitudinal phonons of 5220 m/s \cite{crc}. Calculation of EDCs were done by using the Fowler-Nordheim theory, as in the previous works \cite{yanagisawa09, yanagisawa10, wu08}.

In the simulation, a laser pulse is moved in 0.2 fs steps from -200 fs to 1200 fs across the emission site, where the time zero is defined when the pulse maximum meets the emission site. Note that the time steps were refined until convergence was reached at 0.2 fs. Electron distribution functions and EDCs were calculated at each time step. The resulting transient EDCs were integrated over the entire time interval. EDCs of field emission without laser excitation were also calculated for the rest of one period of the laser pulse repetition cycle (approximately 10 ns), and was added to the EDCs from the first 1400 fs; the resulting time-integrated EDCs were normalized at the maximum intensity. Thus obtained simulations were compared with the measured EDCs normalized at their maximum intensities. Quantitative comparisons based on the absolute intensities were not done because the transmission function of the spectrometer is not well known. There are only three fitting parameters: 1) the work function $\Phi$, 2) the DC field $F_{DC}$, and 3) the excitation constant $C_{exc}$. Fitting was done first by adjusting these three parameters at one particular setting of the tip voltage and laser power ($V_{tip}$=-500 V; $P_{L}$ = 50 mW), and then by simulating time-integrated EDCs for the other settings by scaling up or down $F_{DC}$ and $C_{exc}$ according to the corresponding tip voltage and laser power. This procedure was iterated until reasonable fitting was obtained for all settings.

\begin{table}[t]
\begin{center}
\includegraphics[bb=0 0 552 242, scale=0.42]{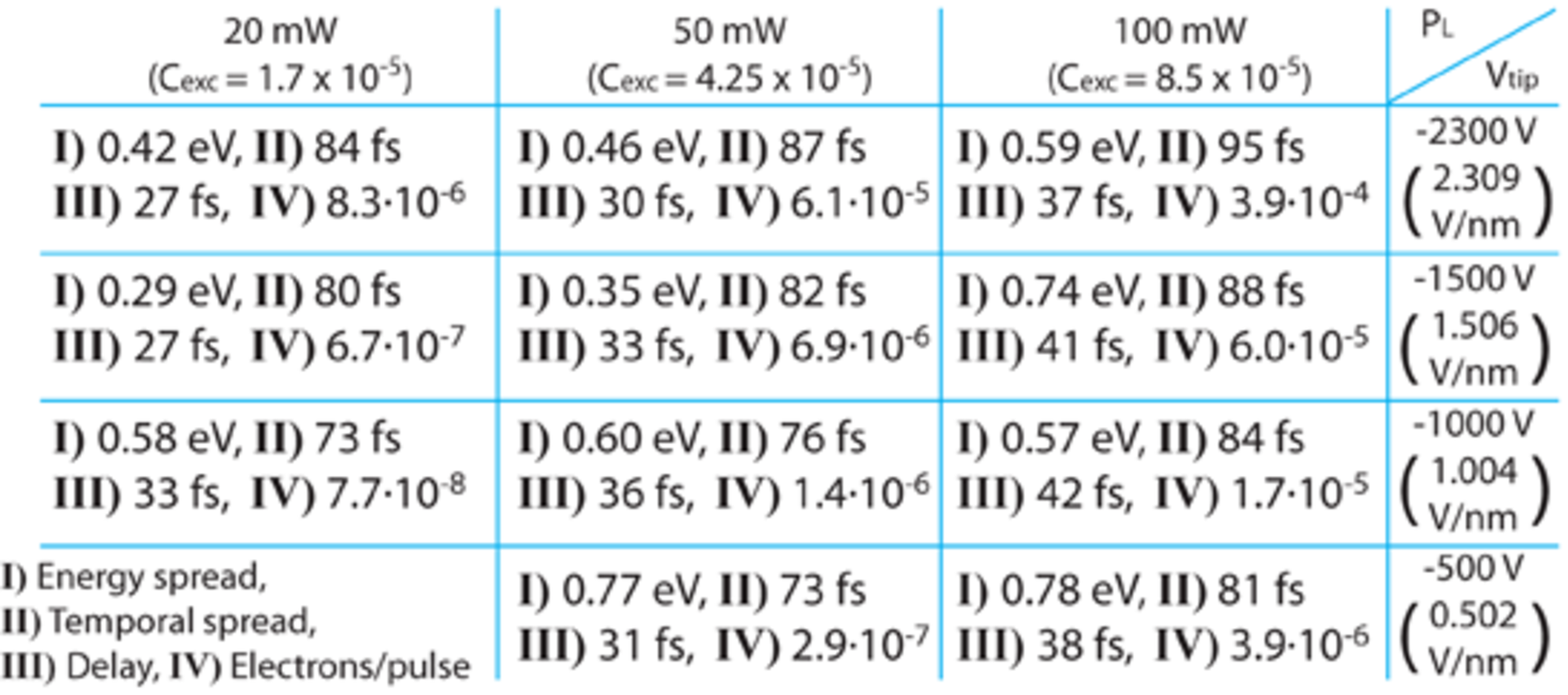}
\end{center}
\vskip -\lastskip \vskip -3pt
\caption{
\label{fig:label-5}
Four characteristic parameters of electron pulses for various tip voltages and laser powers. I) Energy spread, II) temporal spread, III) delay of the electron pulse with respect to the laser pulse and IV) the number of emitted electrons per pulse are listed in this order at each setting. Energy spreads are FWHM of main photo-field emission peaks in measured EDCs in Fig. 3. The temporal spread and delay are taken from temporal profiles of electron pulses, as shown $e.g.$ in Fig. 4(c). The number of electrons per pulse is calculated by assuming that the emission area selected by the pinhole is 10 nm $\times$ 10 nm.}
\vspace{-3pt}
\end{table}

Fig. 4(a) shows examples of the simulations at three time steps. The electron distribution at -100 fs shows a clear multiple step character, but due to the scattering processes the distribution function becomes more and more smeared out as time goes on. Accordingly, the transient EDCs change their shapes and peak positions with time, showing a relaxation of photo-excited electrons to lower energies. The resulting time-integrated EDCs are in good agreement with the corresponding experimental EDCs as shown in Fig. 4(b). From the energy-integrated transient EDCs one can also calculate a temporal line profile of the emitted electron pulse, which is shown in Fig. 4(c). The profile shows a delay of peak emission (31 fs) and squeeze of temporal width (73 fs). The delay is caused by the electron dynamics. The squeeze is due to the nonlinearity of the multi-photon excitations.

All the experimental EDCs in Fig. 3 were reproduced by such simulations. Fitting parameters for $F_{DC}$ and $C_{exc}$ obtained from the calculations shown in Fig. 4(b) were scaled up or down according to the corresponding values of $V_{tip}$ and $P_{L}$. Then the time-integrated EDCs were calculated and are shown in Fig. 3. Throughout all the various conditions, the simulations are in very good accordance with the experiental EDCs. Thus we conclude that electron dynamics play a significant role in photo-field emission.

Finally, we shall comment on the relevance of these findings in view of designing pulsed electron sources for the applications mentioned earlier. From the quantitative description of experimental EDCs achieved with our simulations, we can be confident to predict the energy and temporal spreads of electron pulses, and also the delay of the emission maximum with respect to the laser pulses. As a rule of thumb, these values increase with laser power as shown in Table 1, without even considering space charge effects in the vacuum. Such an increase is not desirable for applications in general. On the other hand, electron emission currents increase with laser power. To reach higher intensities while keeping energy and temporal spreads low, photo-field emission from a tip array \cite{tsujino09,kirk09} with an illumination by low-power laser pulses is considered to be the optimal solution.

This work was supported by the Swiss National Science Foundation through the {\it Ambizione} (grant number PZ00P2\_131701) and the {\it NCCR MUST}. We thank Prof. H. W. Fink for discussion.


\end{document}